\documentclass[prl,twocolumn,showpacs,preprintnumbers,amsmath,amssymb]{revtex4}
\usepackage{amsmath,amssymb}
\usepackage{graphicx}% Include figure files
\usepackage{dcolumn}% Align table columns on decimal point
\usepackage{bm}% bold math

\begin{document}
\title{Optical Lattice Polarization Effects on Hyperpolarizability\\ of Atomic Clock
Transitions}
\author{A. V. Taichenachev and V. I. Yudin}
\affiliation{Institute of Laser Physics SB RAS, Novosibirsk 630090, Russia\\
Novosibirsk State University, Novosibirsk 630090, Russia}
\author{V. D. Ovsiannikov}
\affiliation{Physics Department, Voronezh State University,
Voronezh 394006, Russia}
\author{V. G. Pal'chikov}
\affiliation{Institute of Metrology for Time and Space at National
Research Institute for Physical--Technical and Radiotechnical
Measurements, Mendeleevo, Moscow Region 141579, Russia}
\date{\today}

\begin{abstract}
The light-induced frequency shift due to the hyperpolarizability
(i.e. terms of second-order in intensity) is studied for a
forbidden optical transition, $J$=0$\to$$J$=0. A simple universal
dependence on the field ellipticity is obtained. This result
allows minimization of the second-order light shift with respect
to the field polarization for optical lattices operating at a
magic wavelength (at which the first-order shift vanishes). We
show the possibility for the existence of a magic elliptical
polarization, for which the second-order frequency shift
vanishes. The optimal polarization of the lattice field can be
either linear, circular or magic elliptical. The obtained results
could improve the accuracy of lattice-based atomic clocks.
\end{abstract}

\maketitle

In recent years significant attention has been devoted to optical
lattice atomic clocks \cite{Katori1,Katori2}, in part because the
prospects for a fractional frequency uncertainty of such a clock
could achieve of the level 10$^{-17}$-10$^{-18}$. Apart from
obvious practical applications, such improved clocks will be
critical for a variety of terrestrial and space-borne applications
including improved tests of the basic laws of physics and
searches for drifts in the fundamental constants \cite{Peik}. The
crucial ingredient, for achieving such high metrological
performance, is the existence of the magic wavelength
$\lambda_m$, at which the first-order (in intensity) light shift
of the clock transition $^1$S$_0$$\to$$^3$P$_0$ cancels for
alkaline-earth-like atoms (such as Mg, Ca, Sr, Yb, Zn, Cd). To
date in several experiments cold atoms were trapped in optical
lattices at the magic wavelength and the  clock transition was
observed \cite{Katori2,Ludlow,Bar06,Brusch}. From the
metrological viewpoint even isotopes (with zero nuclear
spin) are more attractive. To excite strictly forbidden clock
transitions in even isotopes the method of magnetic field-induced
spectroscopy was proposed \cite{TY06} and used \cite{Bar06}.

Obviously, the achievement of such an extraordinary accuracy in
frequency standards is a challenging goal. On the way to this
goal it will be necessary to use new approaches and to solve
step-by-step the critical physical problems \cite{Katori2}. For
example, since at the magic wavelength $\lambda_m$ the
first-order shift vanishes, one of the main factors that limits
the accuracy of these optical clocks is the second-order shift
due to the atomic hyperpolarizability. Indeed, for the formation
of optical lattices with the potential depth of order of MHz
\cite{Katori2,Ludlow,Bar06,Brusch} it is necessary to use laser
beams with the intensity at a level of a few tens of kW/cm$^2$.
According to our numerical estimates for different
alkaline-earth-like atoms \cite{Ovs1,Ovs2} and first experimental
observations for Sr \cite{Brusch}, the second-order shift can be
at a level of 1-10 Hz in such high-intensity fields. In this case
to get planned accuracy we need strictly to control the spatially
non-uniform optical lattice fields at a level of
$10^{-3}$-$10^{-5}$ under conditions of strong focusing,
reflections and interference of light beams. Here apart from
long-term stabilization of the laser radiation (power, transverse
distribution of intensity) we need precision long-term stability
of the whole optical system. A significant reduction of a lattice
field intensity is not an effective solution of the problem,
because in this case both the number and lifetime of the trapped
atoms will be reduced also.

Thus, the hyperpolarizability effect on atoms in optical lattices
is an important physical problem, which needs to be carefully
studied (the papers \cite{Katori1,Brusch,Ovs1} have begun such
investigations). In this context the search for alternative
methods of minimization of the second-order shifts is especially
relevant.

In the present paper we study the dependence of the second-order
shift on the elliptical field polarization for the optically
forbidden $J$=0$\to$$J$=0 transition. We show how to minimize
this shift with respect to the ellipticity of the lattice field
polarization. It turns out that under certain conditions there
exists a magic ellipticity at which the second-order light shift
vanishes.

Consider an atom in a monochromatic elliptically polarized light
field with frequency $\omega$:
\begin{equation}\label{E}
{\bf E}(t)={\rm Re}\{E\,{\bf e}\,e^{-i\omega t}\}\,,
\end{equation}
where $E$ is a scalar field amplitude, ${\bf e}$ is a complex unit
polarization vector, $({\bf e}$$\cdot$${\bf e}^*)$=1. If the
quantization axis $Oz$ is orthogonal to the polarization ellipse,
we have the following expansions in Cartesian \{${\bf e}_x$,${\bf
e}_y$,${\bf e}_z$\} and spherical \{${\bf e}_0$=${\bf e}_z$,${\bf
e}_{\pm 1}$=$\mp$(${\bf e}_x$$\pm$$i{\bf e}_y$)/$\sqrt{2}\,$\}
bases:
\begin{eqnarray}\label{e}
&&{\bf e}=\cos(\varepsilon){\bf e}_x+i\sin(\varepsilon){\bf e}_y=
\nonumber \\ &&\quad -\sin(\varepsilon-\pi/4)\,{\bf
e}_{-1}-\cos(\varepsilon-\pi/4)\,{\bf e}_{+1}\,.
\end{eqnarray}
Here the ellipticity angle $\varepsilon$ can take values $-\pi/4
\leq \varepsilon \leq \pi/4$. Obviously (see in Fig.1a),
$|\tan(\varepsilon)|$ is equal to the ratio of the minor axis to
the major axis and the sign of $\varepsilon$ determines the
helicity. Note that  $\varepsilon$=0 corresponds to a linear
polarization, while $\varepsilon$=$\pm\pi$/4 correspond to
circular polarizations. The atom-field interaction will be
considered in the dipole approximation $-$($\hat{\bf d}{\bf E}$).

Each energy level of the forbidden $J_g$=0$\to$$J_e$=0 transition
is shifted by amount $\Delta {\cal E}_j$ ($j$=$e$,$g$) in
response to the field  (\ref{E}). These shifts can be expanded in
series in even powers of the field amplitude $E$:
\begin{equation}\label{DE}
\Delta {\cal E}_j/\hbar =\alpha_j I+\beta_j I^2+...\qquad (j=e,g),
\end{equation}
where  $I$=$c$$|E|^2$/8$\pi$. The first term ($\propto$$I$)
describes a first-order Stark shift, which for levels with  $J$=0
does not depend on the field polarization ${\bf e}$, i.e. the
polarizabilities $\alpha_j$ are completely determined by the
field frequency $\omega$ alone. The second term ($\propto$$I^2$)
in (\ref{DE}) describes energy level shifts due to the
hyperpolarizability (the general expression of the
hyperpolarizability for levels with arbitrary angular momentum is
presented in \cite{DO86}). Coefficients $\beta_j$ depend both on
the frequency $\omega$ and on the ellipticity $\varepsilon$
\cite{MOR76}. From eq.(\ref{DE}) it follows that the frequency of
forbidden transition  $\omega_0$ is shifted by the external field
(\ref{E}) by an amount
\begin{eqnarray}\label{D_omega}
&&\Delta\omega_0\equiv (\Delta {\cal E}_e-\Delta {\cal
E}_g)/\hbar =\widetilde{\alpha}(\omega)I+\widetilde{\beta}(\omega,
\varepsilon)I^2+...\nonumber \\
&&\widetilde{\alpha}(\omega)=\alpha_e-\alpha_g\,,\quad
\widetilde{\beta}(\omega, \varepsilon)=\beta_e-\beta_g\,,\quad
...\,.
\end{eqnarray}
At the magic frequency $\omega_{m}$=2$\pi$$c$/$\lambda_m$ the
first-order Stark shift vanishes, i.e.
$\widetilde{\alpha}(\omega_{m})$=0.

\begin{figure}[htb]
\centerline{\includegraphics[width=8.3cm]{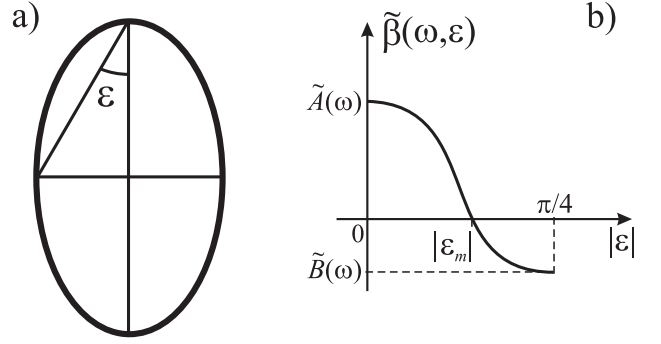}}
\caption{a) Definition of the elliptical polarization parameter $\varepsilon$ (see eq.(\ref{e})).\\
b) Illustration of the existence of a magic elliptical
polarization $\varepsilon_m$ (see eq.(\ref{em})), when the
coefficients $\widetilde{A}(\omega)$ and $\widetilde{B}(\omega)$
have opposite signs.} \label{fig1}
\end{figure}

The dependence of the coefficient
$\widetilde{\beta}$($\omega$,$\varepsilon$) on ellipticity can be
used to minimize the influence of the second term
($\propto$$I^2$). According to electric dipole selection
rules, the second-order shifts for levels with $J$=0 originate
from all possible transitions $J$=0$\to$$J'$=1$\to$$J''$=0,1,2
(quadratic on the field amplitude with frequencies
0,$\pm$2$\omega$), i.e. via virtual levels with $J'$=1. It can be
rigorously proven that each of the three generalized channels
gives only two contributions with different polarization
dependencies in $\beta_j$:
\begin{equation} \label{bt}
\beta_j=\sum_{J''=0,1,2}\left[R^{(j)}_{J''}(\omega)\,{\cal
P}^{}_{J''}({\bf e}) + S^{(j)} _{J''}(\omega)\,{\cal
Q}^{}_{J''}({\bf e})\right],
\end{equation}
where $R^{(j)}_{J''}(\omega)$ and $S^{(j)}_{J''}(\omega)$ depend
on the frequency only. All polarization dependencies are contained
in the factors
\begin{eqnarray} \label{PQ}
{\cal P}^{}_{J''}({\bf e})&=&\left(\{{\bf e}^*\otimes{\bf
e}^*\}^{}_{J''}\cdot \{{\bf e}\otimes{\bf e}\}^{}_{J''}\right),
\nonumber
\\ {\cal Q}^{}_{J''}({\bf e})&=&\left(\{{\bf e}\otimes{\bf e}^*\}^{}_{J''}\cdot \{{\bf e}^*\otimes{\bf
e}\}^{}_{J''}\right),
\end{eqnarray}
which are the scalar products of tensors composed of the unit
polarization vectors ${\bf e}$ and  ${\bf e}^*$ and can be
presented explicitly in terms of the scalar $({\bf e}$$\cdot$$
{\bf e})$ and vector $[{\bf e}$$\times$${\bf e}^*]$ products as
\cite{varsh75}:
\begin{eqnarray}\label{QQPP}
&& {\cal P}^{}_0({\bf e})=|({\bf e}\cdot{\bf
e})|^2/3;\qquad\;\;\; {\cal Q}^{}_0({\bf e})=1/3;\\ && {\cal
P}^{}_1({\bf e})=|[{\bf e}\times{\bf e}]|^2/2\equiv 0;\;\;\,
{\cal Q}^{}_1({\bf e})=|[{\bf e}\times{\bf e}^*]|^2/2;
\nonumber\\
&& {\cal P}^{}_2({\bf e})=1-|({\bf e}\cdot{\bf e})|^2/3;\;\;\;\;
{\cal Q}^{}_2({\bf e})=1/6+|({\bf e}\cdot{\bf e})|^2/2.\nonumber
\end{eqnarray}
Using expansion (\ref{e}), we find:
\begin{equation}\label{ee}
\left|({\bf e}\cdot{\bf e})\right|^2=\cos^2(2\varepsilon)\,,\quad
\left|[{\bf e}\times{\bf e}^*]\right|^2=\sin^2(2\varepsilon)\,.
\end{equation}
Eqs.(\ref{bt}),(\ref{QQPP}) and (\ref{ee}) allow  us to present
the polarization dependence of the coefficients $\beta_j$ in
equation (\ref{DE}) as
$\beta_j$=$A_j$($\omega$)$\,\cos^2$(2$\varepsilon$)+$B_j$($\omega$)$\,\sin^2$(2$\varepsilon$).
As a result, we can write the universal expression for the
polarization dependence of the second-order light shift in the
following simple form:
\begin{equation}\label{beta}
\widetilde{\beta}(\omega,\varepsilon)=\widetilde{A}(\omega)
\cos^2(2\varepsilon)+\widetilde{B}(\omega) \sin^2(2\varepsilon)\,.
\end{equation}
Thus, to reconstruct a complete polarization dependence of the
coefficient $\widetilde{\beta}$($\omega$,$\varepsilon$), we need
to know (by calculation or experiment) its value only in two
points, for example, for linear ($\varepsilon$=0) and circular
($\varepsilon$=$\pm$$\pi$/4) polarizations:
\begin{equation}\label{lc}
\widetilde{A}(\omega)=\widetilde{\beta}(\omega,0)\,,\quad
\widetilde{B}(\omega)=\widetilde{\beta}(\omega,\pm\pi/4)\,.
\end{equation}
It is worth noting, that the terms $R_{J''}(\omega){\cal
P}_{J''}({\bf e})$ in eq.(\ref{bt}) may demonstrate, in
particular, the contributions of two-photon resonances to states
$J''$=0,2. However, the scalar product ${\cal P}_0$(${\bf e}$)
will vanish for circular polarization, so the two-photon
resonances to an excited state $J''$=0 will appear only for the
non-circular polarization, $\varepsilon$$\neq$$\pm\pi$/4 (see eqs.
(\ref{QQPP}) and (\ref{ee})). Two-photon resonances to $J''$=1
states do not appear, as the scalar product ${\cal P}_1$(${\bf
e}$)$\equiv$0 (see eq.(\ref{QQPP})). The terms
$S_{J''}(\omega){\cal Q}_{J''}({\bf e})$ may have large
``resonance'' values when the fine-structure splitting (for
example, between the metastable $^3$P$_0$ and $^3$P$_{1,2}$
sublevels of the $^3$P$_J$ triplet) is small. The matrix elements
with $J''$=$1$ contribute only to the term $S_1(\omega){\cal
Q}_1({\bf e})$ of eq.(\ref{bt}), which disappears for linear
polarization ($\varepsilon$=0), since ${\cal Q}_1({\bf e})$=0 for
${\bf e}$=${\bf e}^*$. Thus, in the vicinity of resonance one can
anticipate a strong dependence of the hyperpolarizability both on
the frequency and polarization of the field.

Eq.(\ref{beta}) allows us to optimize the light shift
(\ref{D_omega}) with respect to the ellipticity parameter
$\varepsilon$. This optimization consists of determining the
optimal ellipticity $\varepsilon_{opt}$, which minimizes the
absolute value $|$$\widetilde{\beta}$($\omega$,$\varepsilon$)$|$,
i.e. $|$$\widetilde{\beta}$($\omega$,$\varepsilon_{opt}$)$|$ =
min\{$|$$\widetilde{\beta}$($\omega$,$\varepsilon$)$|$\}. Such a
minimization is very important for optical frequency standards
based on optical lattices at magic frequency, where the
first-order Stark shifts cancel out
($\widetilde{\alpha}(\omega_{m})$=0) and the higher-order shift
$\propto$$I^2$ becomes one of the main factors limiting  the
accuracy of the future optical frequency standards.

As is seen from (\ref{beta}), if the coefficients
$\widetilde{A}(\omega)$ and $\widetilde{B}(\omega)$ have the same
sign, then the optimal polarization is either linear
($\varepsilon_{opt}$=0) or circular
($\varepsilon_{opt}$=$\pm$$\pi$/4). Apart from this, (\ref{beta})
allows for a very intriguing possibility, when the coefficients
$\widetilde{A}(\omega)$ and $\widetilde{B}(\omega)$ have opposite
signs. In this case a magic elliptical polarization
$\varepsilon_m$ always exists (see in Fig.1b), for which the
second-order light shift vanishes:
\begin{equation}\label{em}
\widetilde{\beta}(\omega,\varepsilon_m)=0\;\Rightarrow\;
\tan(2\varepsilon_m)
=\pm\sqrt{-\widetilde{A}(\omega)/\widetilde{B}(\omega)}
\end{equation}
and, consequently, $\varepsilon_{opt}$=$\varepsilon_m$. Obviously,
the most interesting case arises for the magic ellipticity at the
magic frequency $\omega_m$, i.e. when
$\widetilde{\beta}$($\omega_m$,$\varepsilon_m$)=0. One of possible
candidates for such a remarkable coincidence is Yb.

\begin{figure}[htb]
\centerline{\includegraphics[width=8.3cm]{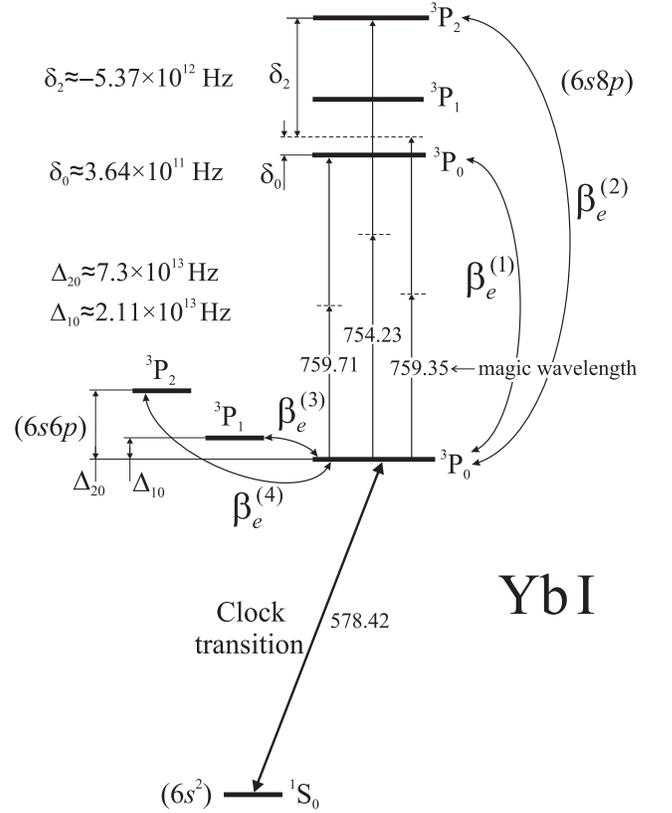}} \caption{Yb
energy levels responsible for the main contributions
$\beta_e^{(1,2,3,4)}$ to the second-order shift of the forbidden
transition (6$s^2$)$^1$S$_0$$\to$(6$s$6$p$)$^3$P$_0$ (all
wavelengths are given in nm).} \label{fig2}
\end{figure}

According to the experimental results \cite{Bar06}, the magic
wavelength $\lambda_m$ for the forbidden transition
(6$s^2$)$^1$S$_0$$\to$(6$s$6$p$)$^3$P$_0$ in Yb equals
approximately to 759.35 nm. Comparing this value with the energy
spectrum, one finds that this wavelength nearly meets the
two-photon resonance conditions that occur at 759.71 nm
((6$s$6$p$)$^3$P$_0$$\to$(6$s$8$p$)$^3$P$_0$ resonance) and
754.23 nm ((6$s$6$p$)$^3$P$_0$$\to$(6$s$8$p$)$^3$P$_2$
resonance). Therefore, the main contribution to the second-order
shift $\propto$$I^2$ is due to the shifts of metastable level
$J_e$=0 (i.e. (6$s$6$p$)$^3$P$_0$), originating from interactions
via levels indicated in Fig.2 (i.e. resonant contributions). Thus,
\begin{equation}\label{b1234}
\widetilde{\beta}(\omega,\varepsilon)\approx \beta_e^{(1)}
+\beta_e^{(2)}+\beta_e^{(3)} +\beta_e^{(4)}\,,
\end{equation}
where terms $\beta_e^{(1,2)}$ are related to the resonance
two-photon transitions
(6$s$6$p$)$^3$P$_0$$\to$(6$s$8$p$)$^3$P$_{0,2}$ at the doubled
magic frequency 2$\omega_m$, and $\beta_e^{(3,4)}$ originate from
the interaction of the level (6$s$6$p$)$^3$P$_0$ with the other
levels (6$s$6$p$)$^3$P$_{1,2}$ of the same fine-structure
manifold. The polarization dependencies for $\beta_e^{(1,2)}$ are
determined by ${\cal P}_{0,2}$(${\bf e}$), whereas for
$\beta_e^{(3,4)}$ are determined by ${\cal Q}_{1,2}$(${\bf e}$).
Note that the resonance two-photon transition $J$=0$\to$$J$=1 is
forbidden in the dipole approximation \cite{Grynberg} (see also
${\cal P}_1$(${\bf e}$)$\equiv$0 in eq.(\ref{QQPP})), therefore
in (\ref{b1234}) the contribution from the transition
(6$s$6$p$)$^3$P$_0$$\to$(6$s$8$p$)$^3$P$_{1}$ (see Fig.2) is
neglected.

Using (\ref{QQPP}) and (\ref{ee}), the terms in (\ref{b1234}) can
be written as:
\begin{eqnarray}\label{beta1}
\beta_e^{(1)}(\omega,\varepsilon)&=&\frac{b_1(\omega)}
{\delta_0}\frac{\cos^2(2\varepsilon)}{3}\geq 0\;\;(\delta_0>0)\\
\label{beta2}
\beta_e^{(2)}(\omega,\varepsilon)&=&\frac{b_2(\omega)}
{\delta_2}\, \frac{3-\cos^2(2\varepsilon)}{3}<0\;\; (\delta_2<0)\\
\label{beta3}
\beta_e^{(3)}(\omega,\varepsilon)&=&-\frac{b_3(\omega)}
{\Delta_{10}}\frac{\sin^2(2\varepsilon)}{2}\leq 0\\ \label{beta4}
\beta_e^{(4)}(\omega,\varepsilon)&=&-\frac{b_4(\omega)}
{\Delta_{20}}\frac{1+3\cos^2(2\varepsilon)}{6} <0\,,
\end{eqnarray}
where the coefficients $b_{1,2,3,4}$ are assumed to be positive,
$\delta_{0}$ and $\delta_{2}$ are the two-photon detunings from
the transitions (6$s$6$p$)$^3$P$_0$$\to$(6$s$8$p$)$^3$P$_{0,2}$,
and $\Delta_{10}$ and $\Delta_{20}$ are the fine-structure
splittings of the (6$s$6$p$)$^3$P$_{0,1,2}$ state (see Fig.2).
From eq.(\ref{beta1}), the term $\beta_e^{(1)}$ is positive
(because $\delta_{0}$$>$0), while all the other terms
$\beta_e^{(2,3,4)}$ are negative.

As it follows from (\ref{beta1}) and (\ref{beta3}), in the case
of linear polarization ($\varepsilon$=0) the term $\beta_e^{(1)}$
becomes  maximal, and $\beta_e^{(3)}$=0. Due to the strong
resonance conditions $|$$\delta_2$/$\delta_0$$|$$\approx$15 and
$\Delta_{20}$/$\delta_0$$\approx$200, we expect that the term
$\beta_e^{(1)}$ will dominate over  $\beta_e^{(2)}$ and
$\beta_e^{(4)}$. This directly leads to a positive value for
$\widetilde{\beta}$($\omega_m$,0), i.e.
$\widetilde{A}$($\omega_m$)$>$0 in accordance with (\ref{lc}).
For circular polarization ($\varepsilon$=$\pm$$\pi$/4) we have
$\beta_e^{(1)}$=0 and $\beta_e^{(2,3,4)}$$<$0, which leads to a
negative $\widetilde{\beta}$($\omega_m$,$\pm$$\pi$/4), i.e.
$\widetilde{B}$($\omega_m$)$<$0 according to (\ref{lc}). Thus,
the coefficients $\widetilde{A}(\omega_m)$ and
$\widetilde{B}(\omega_m)$ may have opposite signs, thus providing
a sufficient condition for the existence of a magic elliptical
polarization $\varepsilon_m$ (\ref{em}) at the magic frequency
$\omega_m$ for Yb.

It should be stressed that the qualitative analysis above does
not guarantee the existence of a magic ellipticity for Yb,
because we did not take into account contributions to
hyperpolarizability from the inner-shell electrons, nor numerous
off-resonant contributions in
$\widetilde{\beta}$($\omega_m$,$\varepsilon$) of the jumping
electron. So, while there is a good chance of a magic ellipticity
for Yb, the ultimate answer will be given by an experiment.
Nevertheless, the analysis shows that the presence of the
near-resonance two-photon transitions can lead to an intriguing
situation. In this context, it is worth noting that for Sr atoms
there also is a near-resonant two-photon transition
(5$s$5$p$)$^3$P$_0$$\to$(5$s$4$f$)$^3$F$_2$ \cite{Brusch}. In
\cite{Brusch} the second-order shifts have been investigated only
for a linearly polarized field. From the experimental results it
follows that $\widetilde{A}$($\omega_m$)$>$0. The value and sign
of second-order shift in circularly polarized field (i.e. the
coefficient $\widetilde{B}$($\omega_m$)) are still unknown.
Moreover, for circular polarization the negative contribution due
to the interaction with the level (5$s$5$p$)$^3$P$_1$ (analogue
of the term $\beta_e^{(3)}$ in Fig.2) becomes maximal (see
eq.(\ref{beta3})), while for linear polarization it equals to
zero. Consequently, the question of the optimal polarization
remains open and the possibility of a magic ellipticity for Sr
still takes play.

Concluding, for forbidden optical transition $J$=0$\to$$J$=0 (for example,
$^1$S$_0$$\to$$^3$P$_0$ clock transition in even isotopes of alkaline-earth-like atoms) we
have investigated the polarization dependence of the higher-order
frequency shifts $\propto$$I^2$, originating from the atomic
hyperpolarizability. This dependence has a simple universal form
(\ref{beta}) and we have described the method for minimizing the
second-order shift for optical lattices at the magic frequency
$\omega_m$. To this end, the higher-order shifts for linear and
circular field polarization should be measured and compared. If
these shifts are the same sign, then the optimal polarization
(either circular or linear) will correspond to minimal absolute
value of the shift. If the signs of the measured shifts are
different, then a magic ellipticity $\varepsilon_m$ will exist,
where the second-order shift vanishes. The magic ellipticity can
be estimated from (\ref{em}), and determined more accurately from
experiments.

It should be stressed that the existence of a magic ellipticity
allows a practically ideal one-dimensional standing wave optical
lattices for the frequency standards, because in this case it is
not necessary to control strictly the lattice field intensity.
Consequently, one can use high-intensity fields to create deep
potential lattices with high efficiency of trapping and with
longer capture time. Note also, that in deep potential lattices
cold atoms are localized on length scales much less than the field
wavelength, i.e. the strong Lamb-Dicke regime is realized.

These results can be extended to 2D and 3D lattices in the field
with spatially non-uniform polarization \cite{Jessen}. Here the
lattice field configuration should be chosen in such a way that
at the potential energy minimum the local field polarization
coincides with the optimal value (either linear, circular or
magic) for the given element. In this case for lower vibrational
levels the second-order shifts will be minimal, assuming the
Lamb-Dicke regime.

We thank C. W. Oates, C. W. Hoyt, Z. W. Barber, and L. Hollberg
for helpful discussions. AVT and VIYu were supported by RFBR
(grants 05-02-17086, 04-02-16488, 05-08-01389), VDO acknowledges
the support from the CRDF (USA) and MinES of Russia (BRHE program,
award VZ-010).\\
AVT and VIYu e-mail address: llf@laser.nsc.ru

\end{document}